\documentclass[10pt]{article}
\usepackage{amsmath}
\usepackage{booktabs}
\usepackage{fancyhdr}
\usepackage[dvipsnames]{xcolor}
\usepackage[most]{tcolorbox}

\usepackage{pifont} 
\newcommand{\cmark}{\ding{51}}
\usepackage{dirtree}

\usepackage[round]{natbib}
\usepackage{geometry}
\usepackage{graphicx}

\usepackage{charter}
\usepackage{hyperref}
\hypersetup{colorlinks=true,allcolors=BlueViolet}

\title{\textsc{\textbf{Towards an open registry of\\Earth observation instruments}}}

\author{
David Montero\textsuperscript{1,*}\quad 
César Aybar\textsuperscript{2}\quad 
Miguel D. Mahecha\textsuperscript{1,3}\quad
Luis Gómez-Chova\textsuperscript{2} \\
\small\textsuperscript{1}IEF, Leipzig University\quad
\textsuperscript{2}IPL, Universitat de Val\`encia\quad
\textsuperscript{3}UFZ\\
\small*Corresponding author: \texttt{david.montero@uni-leipzig.de}
}
\date{}

\begin{document}

\maketitle

\begin{abstract}
Earth observation (EO) is essential to understanding the Earth system, enabling the transformation of planetary properties into measurable variables that can be analysed, compared, and modelled. In recent decades, EO capabilities have grown rapidly, accompanied by an even faster expansion in the number and variety of available EO instruments.
Today, EO includes instruments deployed on satellites, airborne platforms, and terrestrial or in-situ systems. However, despite this proliferation of instruments, users often lack a single, reliable source describing their existence and key characteristics.
Although existing data catalogues have substantially improved dataset discovery, they primarily describe data products rather than providing persistent, curated metadata about the instruments that produced them.
Here we present Awesome Earth Observation Instruments, an open, standardized, and community-oriented registry providing machine-readable metadata for EO instruments. The catalogue is hosted on GitHub and allows contributors to submit instrument metadata following a common schema. The schema combines a lightweight core with modular extensions covering spectral, geometric, and data access-related metadata, enabling both standardization and flexibility across diverse EO systems.
All submissions undergo automated schema validation and human review. Because the schema is open, versioned, and extensible, the catalogue can continuously evolve as new instruments and metadata requirements emerge. This facilitates the discovery, interpretation, and analysis of EO data in light of instrument characteristics.
We anticipate that the catalogue will serve as a community reference for EO instrument metadata while lowering barriers to discovery and integration. To support programmatic access and interoperability, we further envisage an API for integration within common EO analysis environments. The catalogue is openly available at \url{https://github.com/awesome-spectral-indices/awesome-earth-observation-instruments}.
\end{abstract}

\thispagestyle{fancy}
\renewcommand{\headrulewidth}{0pt}
\renewcommand{\footrulewidth}{0pt}
\fancyhf{}

\section{Introduction}\label{MANUSCRIPT}
 
\sloppy

Earth observation (EO) through remote sensing is fundamental for understanding the Earth system because it provides repeated, spatially explicit measurements that transform planetary properties into variables that can be analysed, compared, and modelled \citep{running1994terrestrial,justice1998moderate}. As the demand for such observations has grown, EO capabilities have expanded rapidly over recent decades, accompanied by a substantial increase in the number and diversity of remote sensing instruments available to the community \citep{wulder2012opening}. Contemporary EO includes instruments onboard satellite platforms, airborne sensors, and terrestrial or in-situ systems. Passive optical instruments measure reflected radiation using multispectral \citep{wulder2016global,drusch2012sentinel2} or hyperspectral systems \citep{clark2024imaging}. These measurements commonly cover the ultraviolet, visible, near-infrared, and short-wave infrared regions, with some instruments extending into the thermal infrared. Active systems, such as synthetic aperture radar, provide illumination-independent observations and enable monitoring under clouds and variable atmospheric conditions \citep{torres2012gmes}. Airborne and uncrewed aerial vehicle (UAV) instruments can provide centimetre-scale spatial detail and flexible acquisition timing, while ground-based sensors offer direct in-situ reference measurements \citep{alvarez2021uav}.

This diversity of platforms and sensing modalities creates a practical challenge: users often lack a single reliable source for identifying which instruments exist and for accessing their key characteristics, including governance, mission status, spectral configuration, imaging geometry, spatio-temporal resolution, and data access points. Existing resources only partially address this need \citep{bugbee2021improving}. Some instrument catalogues, such as the sensor information provided by the Index DataBase \citep{henrich2009onlineindices}, include useful entries but are limited in openness, completeness, and update frequency. The Committee on Earth Observation Satellites (CEOS) Database\footnote{{https://database.eohandbook.com}} provides one of the most comprehensive and standardised overviews of EO missions and instruments, but it does not systematically expose the level of instrument-specific detail required for many analytical workflows, such as spectral characteristics or data-access patterns. In parallel, data catalogues, including large data platforms and SpatioTemporal Asset Catalog (STAC)\footnote{{https://stacspec.org}} catalogues, have substantially improved dataset discovery. However, they are primarily designed to describe data-product collections and associated assets, rather than to provide persistent, curated descriptions of the instruments that generated those data products.

This paper presents the first iteration of Awesome Earth Observation Instruments, an open, standardised, and community-oriented registry of EO instruments that provides machine-readable instrument metadata. The registry is designed to complement existing data catalogues and to support reproducible, automated geospatial workflows. It is implemented as a living catalogue hosted on GitHub, where contributors can add instruments and associated metadata according to a shared specification. The specification is designed to be straightforward to read and implement while remaining strict where standardisation is essential. Contributions are reviewed by maintainers and validated against a YAML-based JSON Schema. The schema consists of a lightweight core and a set of modular extensions, enabling the catalogue to represent diverse EO systems while remaining extensible. In version 0.1.0, the registry includes 16 catalogue entries covering seven instrument families from widely used satellite EO systems: the Landsat archive (MSS, TM, ETM+, OLI, and TIRS), the Sentinel-2 MultiSpectral Instrument (MSI), and the hyperspectral EMIT instrument on board the International Space Station (ISS).

The remainder of the paper is organised as follows. Section~\ref{catalogue} presents the Awesome Earth Observation Instruments catalogue, including the core schema, extensions, repository structure, validation and catalogue-generation workflow, and release and distribution strategy. Section~\ref{showcase} presents a showcase in which an initial set of EO instruments was validated and released in the first catalogue version. Section~\ref{discussion} discusses the potential integration of the catalogue with existing resources, its relevance for artificial intelligence (AI) workflows, possible future extensions, programmatic access, and current limitations. Section~\ref{conclusions} concludes the paper.

\section{The Awesome Earth Observation Instruments Catalogue}\label{catalogue}

The Awesome Earth Observation Instruments Catalogue (AEOI) is a curated, machine-readable, and community-driven registry of EO instruments. It is designed to describe instruments deployed on diverse platform types, including terrestrial systems, airborne and UAV platforms, and spaceborne missions. The catalogue is hosted on GitHub and distributed in JavaScript Object Notation (JSON) format through GitHub and Zenodo. Each registry item corresponds to an EO instrument and is contributed through GitHub according to a predefined schema. To balance human readability with machine interoperability, all input files, including the schema and its extensions, are written in YAML (Sections~\ref{schema}~and~\ref{extensions}).

\subsection{Schema}\label{schema}

The registry schema defines how each instrument item must be organised and is implemented using JSON Schema draft 2020-12\footnote{https://json-schema.org/draft/2020-12}. It is provided as a core YAML file that can be extended through references to additional schemas (Section~\ref{extensions}). The core schema captures general instrument metadata, including identifying attributes such as name, acronym, type, and platform, among others (Table~\ref{tab:instrument_schema}). These attributes are required for ingestion into the catalogue. Each instrument must also provide a unique identifier through the \texttt{id} attribute. We recommend combining the instrument acronym and platform identifier, for example \texttt{MSI\_S2A} for the MultiSpectral Instrument onboard Sentinel-2A. The same identifier must be used as the filename of the corresponding YAML item.

\begin{table}[t]
\footnotesize
\centering
\setlength{\tabcolsep}{4pt}
\renewcommand{\arraystretch}{1.15}
\begin{tabular}{l c l p{9.5cm}}
\toprule
\textbf{Property} & \textbf{Required} & \textbf{Type} & \textbf{Description} \\
\midrule

id             & \cmark & string & Unique alphanumerical identifier of the instrument \\ \midrule
name           & \cmark & string & Full instrument name \\ \midrule
acronym        & \cmark & string & Short instrument acronym \\ \midrule
type           & \cmark & string & Sensing modality; one of: \texttt{multispectral}, \texttt{hyperspectral}, \texttt{radar}, \texttt{rgb}, \texttt{lidar}, \texttt{other} \\ \midrule
platform\_type & \cmark & string & Platform class; one of: \texttt{satellite}, \texttt{airborne}, \texttt{uav}, \texttt{terrestrial} \\ \midrule
platform       & \cmark & array  & List of platform names (e.g., satellite mission, UAV model, aircraft) \\ \midrule
operator       & \cmark & array  & List of operating organizations; predefined (e.g., \texttt{ESA}, \texttt{NASA}, \texttt{JAXA}, \texttt{NOAA}, \texttt{DLR}, \texttt{ISRO}, \texttt{Copernicus}) or custom strings for non-predefined organizations \\ \midrule
start\_date    & \cmark & string & Start of operation (YYYY-MM-DD) \\ \midrule
status         & \cmark & string & Lifecycle state; one of: \texttt{operational}, \texttt{retired}, \texttt{experimental}, \texttt{planned} \\ \midrule
availability   & \cmark & string & Data access level; one of: \texttt{public} or \texttt{private} \\ \midrule
end\_date      &        & string & End of operation (YYYY-MM-DD), if applicable \\ \midrule
notes          &        & string & Free-form additional information \\ \midrule
references     & \cmark & array  & List of reference URLs (papers, documentation, web pages) \\ \midrule
data\_links    &        & array  & URLs providing access to data products \\ \midrule
extensions     &        & object & Optional structured extensions: \texttt{spectral} (band/SRF info; see Table~\ref{tab:spectral_metadata}), \texttt{imaging} (FOV, IFOV, optics; see Table~\ref{tab:geometric_metadata}), \texttt{ee} (Google Earth Engine dataset metadata; see Table~\ref{tab:ee_metadata}), \texttt{planetary\_computer} (Planetary Computer dataset metadata; see Table~\ref{tab:pc_metadata}) \\

\bottomrule
\end{tabular}
\caption{Schema definition for instrument metadata.}
\label{tab:instrument_schema}
\end{table}

\subsection{Extensions}\label{extensions}

EO instruments differ substantially in their mission objectives, platform constraints, sensing principles, and available metadata. Including all possible properties in the core schema would make the catalogue unnecessarily complex and difficult to maintain. Instead, following the general logic of the SpatioTemporal Asset Catalog (STAC) extension mechanism, AEOI uses a compact core schema complemented by optional extensions. The core schema includes an \texttt{extensions} property that can reference additional schemas. When an extension is used for a given instrument, the corresponding properties are validated according to the referenced schema.

The initial implementation includes four extensions: \texttt{spectral}, describing spectral characteristics such as bands and spectral response functions; \texttt{imaging}, describing optical and geometric properties; \texttt{ee}, describing Google Earth Engine (GEE) data access; and \texttt{planetary\_computer}, describing Microsoft Planetary Computer data access.

\subsubsection{Spectral:}\label{ext:spectral}

The spectral extension specifies the spectral characteristics of an instrument (Table~\ref{tab:spectral_metadata}). Although optional, it represents one of the most important metadata components for many EO applications and is therefore strongly recommended whenever the required information is available. The extension remains optional to accommodate instruments for which spectral characteristics are unavailable, variable, or not applicable, such as some RGB cameras, integrated UAV cameras, and flexible systems such as spectroradiometers. The extension can be added later if reliable spectral information becomes available.

The extension covers two main metadata components: spectral bands and spectral response functions (SRFs). For multispectral instruments, spectral bands can be defined either as an inline dictionary of band properties (the \texttt{bands} property in Table~\ref{tab:spectral_metadata}) or as a CSV file containing the same information in tabular form. For hyperspectral instruments, bands can alternatively be generated from a continuous spectral range (the \texttt{range} property in Table~\ref{tab:spectral_metadata}). For a given instrument, only one of these two alternatives must be provided. SRFs are represented as CSV files for simplicity and interoperability. In these files, the first column contains wavelength values and subsequent columns correspond to the identifiers of previously defined spectral bands; cell values represent relative response at each wavelength. Although CSV filenames are explicitly declared in the schema and therefore need not follow a fixed convention, we recommend using the instrument identifier followed by the suffix \texttt{\_SRF} or \texttt{\_BANDS}, depending on the file content (e.g. \texttt{ETM\_L7\_BANDS.csv} for the bands file of the Landsat~7 ETM+ instrument).

\begin{table}[t]
\footnotesize
\centering
\setlength{\tabcolsep}{4pt}
\renewcommand{\arraystretch}{1.15}
\begin{tabular}{l l p{9.5cm}}
\toprule
\textbf{Property} & \textbf{Type} & \textbf{Description} \\
\midrule

bands  & oneOf  & Either (i) CSV filename, or (ii) dictionary of bands keyed by ID (e.g., \texttt{B1}). Each band requires \texttt{center\_wavelength} (nm) and \texttt{bandwidth} (nm), and may include \texttt{band\_description}, \texttt{common\_name} (e.g., \texttt{blue}, \texttt{red}, \texttt{nir}, \texttt{swir}), \texttt{gsd} (m), and \texttt{snr} \\ \midrule

range  & object & Continuous spectral coverage for sensors without discrete bands: \texttt{min} and \texttt{max} wavelength (nm, required), plus either \texttt{sampling} (nm) or \texttt{total\_bands} \\ \midrule

spectral\_response\_function & string & Filename of the spectral response function (SRF) CSV file \\

\bottomrule
\end{tabular}
\caption{Metadata fields for the Spectral Extension (either \texttt{bands} or \texttt{range} must be provided).}
\label{tab:spectral_metadata}
\end{table}

\subsubsection{Imaging:}\label{ext:imaging}

The imaging extension specifies optical and geometric instrument characteristics (Table~\ref{tab:geometric_metadata}). It is optional and can be added whenever the relevant information is available; all properties in the extension are also optional. The extension includes properties commonly used for satellite instruments, such as \texttt{swath\_width}, \texttt{across\_fov}, and \texttt{along\_fov}, as well as properties relevant to airborne and UAV systems, such as \texttt{hfov} and \texttt{vfov}. Unlike the spectral extension, the imaging extension does not require auxiliary files and is populated directly through numerical values.

\begin{table*}[t]
\footnotesize
\centering
\setlength{\tabcolsep}{4pt}
\renewcommand{\arraystretch}{1.15}
\begin{tabular}{l l p{10.5cm}}
\toprule
\textbf{Property} & \textbf{Type} & \textbf{Description} \\
\midrule

swath\_width     & number & Swath width (km) \\ \midrule

across\_fov      & number & Across-track Field of View (FOV) (degrees). Relevant for satellite platforms \\ \midrule

along\_fov       & number & Along-track Field of View (FOV) (degrees). Relevant for satellite platforms \\ \midrule

across\_ifov     & number & Across-track Instantaneous Field of View (IFOV) (microradians). Relevant for satellite platforms \\ \midrule

along\_ifov      & number & Along-track Instantaneous Field of View (IFOV) (microradians). Relevant for satellite platforms \\ \midrule

hfov             & number & Horizontal Field of View (HFOV) (degrees). Relevant for UAV or similar platforms \\ \midrule

vfov             & number & Vertical Field of View (VFOV) (degrees). Relevant for UAV or similar platforms \\ \midrule

entrance\_pupil  & number & Diameter of the entrance pupil (mm) \\ \midrule

focal\_length    & number & Focal length (mm) \\ \midrule

fnumber          & number & F-number \\ \midrule

gsd              & number & Ground Sampling Distance (m). If GSD varies by band, it can be defined per band in the spectral bands CSV file \\

\bottomrule
\end{tabular}
\caption{Metadata fields for the Imaging Extension.}
\label{tab:geometric_metadata}
\end{table*}

\subsubsection{Earth Engine:}\label{ext:ee}

Google Earth Engine (GEE) is a geospatial platform provided by Google that combines cloud-computing resources with a large catalogue of raster and vector geospatial datasets, including many EO products \citep{gorelick2017gee}. To support open data discovery and open-science workflows, the GEE extension provides the first data-access extension in AEOI (Table~\ref{tab:ee_metadata}). Although the core schema includes a generic \texttt{data\_links} property, specialised data-access extensions provide standardised entry points to datasets produced by a given instrument. The extension defines a required \texttt{primary} data-access point for the instrument or platform, for which we recommend bottom-of-atmosphere (BOA) data when available. It also supports optional raw, BOA, and top-of-atmosphere (TOA) data-access points. Each property is an object containing a link to GEE documentation through \texttt{docs} and the corresponding collection identifier in the GEE catalogue. This structure enables users and software tools to retrieve data-access entry points programmatically in a consistent manner.

\begin{table*}[t]
\footnotesize
\centering
\setlength{\tabcolsep}{4pt}
\renewcommand{\arraystretch}{1.15}
\begin{tabular}{l c l p{10cm}}
\toprule
\textbf{Property} & \textbf{Required} & \textbf{Type} & \textbf{Description} \\
\midrule

primary & \cmark & object & Primary dataset; requires \texttt{docs} (documentation URL) and \texttt{collection} (Earth Engine collection ID) \\ \midrule

raw     &        & object & Optional raw (unprocessed) dataset; same structure as \texttt{primary} (\texttt{docs}, \texttt{collection}) \\ \midrule

boa     &        & object & Optional bottom-of-atmosphere (BOA) dataset; same structure as \texttt{primary} (\texttt{docs}, \texttt{collection}) \\ \midrule

toa     &        & object & Optional top-of-atmosphere (TOA) dataset; same structure as \texttt{primary} (\texttt{docs}, \texttt{collection}) \\

\bottomrule
\end{tabular}
\caption{Metadata fields for the Earth Engine Extension.}
\label{tab:ee_metadata}
\end{table*}

\subsubsection{Planetary Computer:}\label{ext:pc}

Microsoft Planetary Computer also provides access to an extensive EO data catalogue, but organises its collections through STAC. Because it is a widely used platform for EO data discovery and retrieval, we implemented a dedicated Planetary Computer extension following the same general data-access pattern used for GEE (Table~\ref{tab:pc_metadata}). The extension adds a required \texttt{stac\_endpoint} property, fixed to the Planetary Computer STAC API endpoint. The remaining properties are equivalent to those in the GEE extension: documentation links should point to the corresponding Planetary Computer dataset documentation, and collection values should match the appropriate STAC collection identifiers.

\begin{table}[t]
\footnotesize
\centering
\setlength{\tabcolsep}{4pt}
\renewcommand{\arraystretch}{1.15}
\begin{tabular}{l c l p{9.5cm}}
\toprule
\textbf{Property} & \textbf{Required} & \textbf{Type} & \textbf{Description} \\
\midrule

stac\_endpoint & \cmark & string & STAC API endpoint (fixed: \url{https://planetarycomputer.microsoft.com/api/stac/v1}) \\ \midrule

primary        & \cmark & object & Primary dataset; requires \texttt{docs} (documentation URL) and \texttt{collection} (STAC collection ID) \\ \midrule

raw            &        & object & Optional raw (unprocessed) dataset; same structure as \texttt{primary} (\texttt{docs}, \texttt{collection}) \\ \midrule

boa            &        & object & Optional bottom-of-atmosphere (BOA) dataset; same structure as \texttt{primary} (\texttt{docs}, \texttt{collection}) \\ \midrule

toa            &        & object & Optional top-of-atmosphere (TOA) dataset; same structure as \texttt{primary} (\texttt{docs}, \texttt{collection}) \\

\bottomrule
\end{tabular}
\caption{Metadata fields for the Planetary Computer Extension.}
\label{tab:pc_metadata}
\end{table}

\subsection{Repository Structure}

\begin{figure}[ht!]
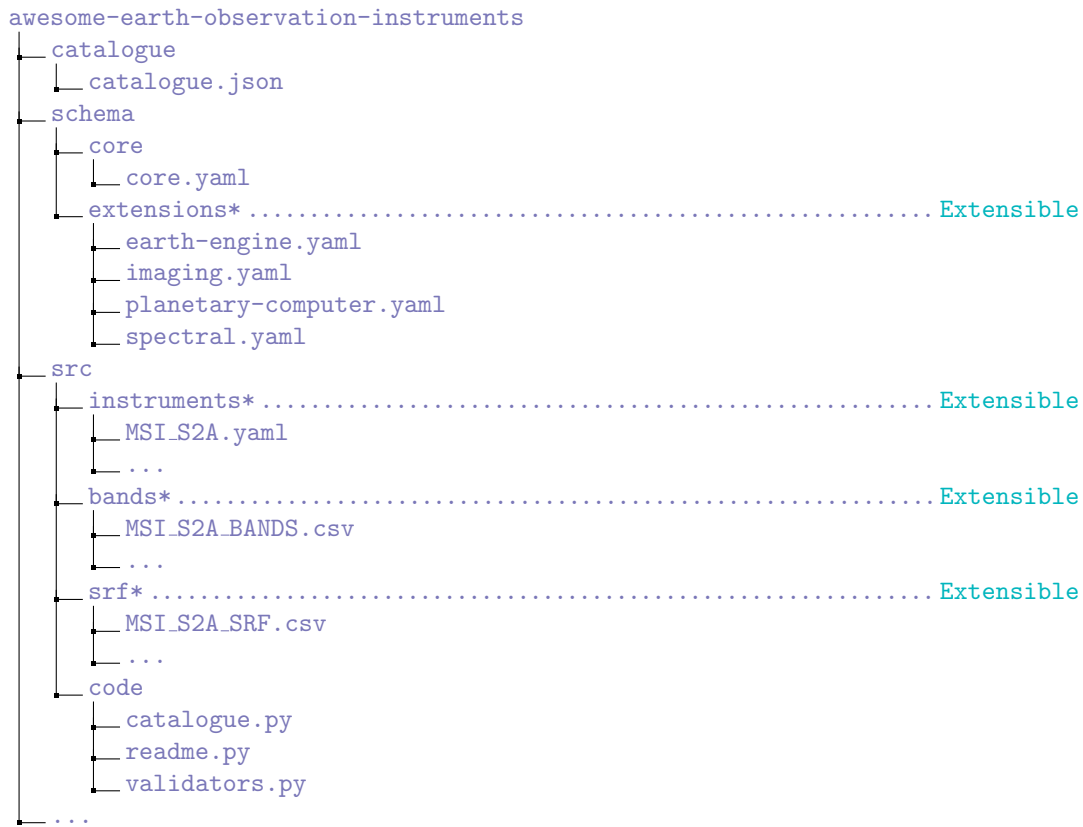

\renewcommand*\DTstylecomment{\ttfamily\color{Aquamarine}}
\renewcommand*\DTstyle{\ttfamily\textcolor{Periwinkle}}

\dirtree{%
    .1 awesome-earth-observation-instruments.
    .2 catalogue.
    .3 catalogue.json.
    .2 schema.
    .3 core.
    .4 core.yaml.
    .3 extensions*\DTcomment{Extensible}.
    .4 earth-engine.yaml.
    .4 imaging.yaml.
    .4 planetary-computer.yaml.
    .4 spectral.yaml.
    .2 src.
    .3 instruments*\DTcomment{Extensible}.
    .4 MSI\_S2A.yaml.
    .4 \dots.
    .3 bands*\DTcomment{Extensible}.
    .4 MSI\_S2A\_BANDS.csv.
    .4 \dots.
    .3 srf*\DTcomment{Extensible}.
    .4 MSI\_S2A\_SRF.csv.
    .4 \dots.
    .3 code.
    .4 catalogue.py.
    .4 readme.py.
    .4 validators.py.
    .2 \dots.
    }

\begin{center}
	
	\caption{\textbf{Repository structure.} Only the folders \texttt{catalogue}, \texttt{schema}, and \texttt{src} are shown here. Folders marked with an asterisk indicate locations where the community can contribute new files.}
\label{fig:repo}
\end{center}
\end{figure}

The repository is organised to facilitate community contributions to the catalogue (Figure~\ref{fig:repo}). The root repository is divided into four main folders: \texttt{catalogue}, \texttt{docs}, \texttt{schema}, and \texttt{src}.

\begin{itemize}

    \item \textbf{\texttt{catalogue}:} This folder contains the generated catalogue JSON file created after audit and validation. Its content is generated automatically.

    \item \textbf{\texttt{docs}:} This folder contains files used to generate the \texttt{README}. It can be extended in the future to support website deployment.

    \item \textbf{\texttt{schema}:} This folder contains the catalogue core schema and its extensions, stored in the \texttt{core} and \texttt{extensions} subfolders, respectively. The \texttt{core} folder contains only the core YAML schema (Section~\ref{schema}). The core schema can be modified by maintainers or through community pull requests after maintainer review, for example when adding new core properties or registering new extensions. The \texttt{extensions} folder contains the YAML schemas used to extend the core schema. New extension schemas can be contributed through pull requests and must be referenced from the core schema.

    \item \textbf{\texttt{src}:} This folder contains all input files required to generate the catalogue and is organised into four subfolders:

    \begin{itemize}
    
        \item \textbf{\texttt{instruments}:} This folder contains the YAML files corresponding to each instrument added to the catalogue. These files must comply with the core schema and with any extensions they use. Users can contribute new instruments through pull requests.

        \item \textbf{\texttt{bands}:} This folder contains CSV files describing spectral bands associated with specific instruments. If the spectral extension is used and the \texttt{bands} property points to a CSV file, that file must be stored in this folder.

        \item \textbf{\texttt{srf}:} This folder contains CSV files describing the spectral response function (SRF) associated with specific instruments. If the spectral extension is used and the \texttt{spectral\_response\_function} property points to a CSV file, that file must be stored in this folder.

        \item \textbf{\texttt{code}:} This folder contains the Python code used to validate catalogue items (\texttt{validators.py}), generate the catalogue (\texttt{catalogue.py}), and generate the repository \texttt{README.md} file (\texttt{readme.py}).
        
    \end{itemize}
    
\end{itemize}

\subsection{Instruments Validation and Catalogue Generation}

Catalogue generation requires a complete validation process for each instrument provided as a YAML input file. The validation workflow has two phases: a human audit performed by repository maintainers and automatic validation of the input data. Instruments are included in the generated catalogue only after both phases are completed successfully.

\subsubsection{Human Audit}

The human audit is performed by a maintainer of the GitHub repository, following a procedure similar to that used in the Awesome Spectral Indices (ASI) catalogue. During this phase, the maintainer checks whether the information provided in the instrument YAML file is supported by the sources listed in the \texttt{references} property. A revision is requested if the references do not point to authoritative or trustworthy sources, such as the organisation responsible for the instrument or its operation. A revision is also requested when the referenced sources are reliable but the submitted metadata do not match the information reported in them.

\subsubsection{Automatic Validation}

After the human audit, the YAML file is automatically validated using a JSON Schema draft 2020-12 validator. This step verifies that the file and its property values comply with the core schema and with any declared extensions. If validation fails, the file must be corrected before the instrument can be included in the catalogue. Maintainers may correct errors when sufficient information is available in the submitted file and references; otherwise, a revision is requested from the contributor. If the YAML file uses the spectral extension and provides CSV files for either the \texttt{bands} or \texttt{spectral\_response\_function} properties, these files are also validated automatically. For \texttt{bands} files, validation fails if required columns are missing or if unsupported columns are present. For SRF files, validation fails if the wavelength column or the columns corresponding to band identifiers defined in \texttt{bands} are missing. As in the YAML validation step, maintainers may correct issues when sufficient information is available, otherwise a revision is requested.

\subsubsection{Catalogue Generation}

After all instruments pass validation, the catalogue is generated as a single JSON file using \texttt{catalogue.py}. During generation, each YAML file is read and validated with helper functions from \texttt{validators.py}. If CSV files are provided for the \texttt{bands} or \texttt{spectral\_response\_function} properties, they are converted into dictionary objects and nested under the corresponding property. If bands are specified through a spectral range, band properties such as centre wavelength and bandwidth are generated automatically from the minimum wavelength, maximum wavelength, and total number of bands.

The final JSON catalogue contains four top-level properties: 1) \texttt{name}, the catalogue name, Awesome Earth Observation Instruments; 2) \texttt{version}, the catalogue version; 3) \texttt{link}, the URL of the JSON file in the GitHub repository; and 4) \texttt{instruments}, a key--value object in which each key is an instrument identifier and each value contains the validated metadata for that instrument.

\subsection{Release and Distribution}

The catalogue is distributed through GitHub releases, with each release version matching the catalogue version. The repository is also linked to Zenodo, enabling archival releases with a dedicated DOI for each version in addition to a general DOI for the repository.

\section{Showcase: Pilot Instruments}\label{showcase}

To evaluate the catalogue implementation and the behaviour of the specified schema, we ingested 16 YAML files corresponding to seven instrument families (MSS, TM, ETM+, OLI, TIRS, MSI, and EMIT) across multiple platforms. Separate catalogue entries are needed when an instrument family has platform-specific configurations or operational histories. For example, the MSI instruments onboard Sentinel-2A, Sentinel-2B, and Sentinel-2C have platform-specific characteristics and operational dates, and are therefore represented as separate items. All submitted files complied with the specified schema.

These instruments were selected because of their relevance to existing resources such as Awesome Spectral Indices (ASI) and the Transparent Access to Cloud-Optimized datasets (TACO) specification. They also cover a broad range of characteristics useful for testing the schema, including multispectral and hyperspectral systems, retired and operational missions, and datasets available through Earth Engine or Planetary Computer. This initial set demonstrated the functionality of both the core schema and the four implemented extensions: spectral, imaging, Earth Engine, and Planetary Computer.

Given the importance of the spectral extension, we ingested CSV files describing the band specifications for all multispectral instruments and, when available, corresponding SRF CSV files. For EMIT, the hyperspectral test instrument, band information was generated from a spectral range defined by minimum wavelength, maximum wavelength, and number of bands. In both cases, ingestion was successful and the catalogue was correctly validated and generated.

\begin{figure}[ht!]
\begin{center}
	\includegraphics[width=1.0\textwidth]{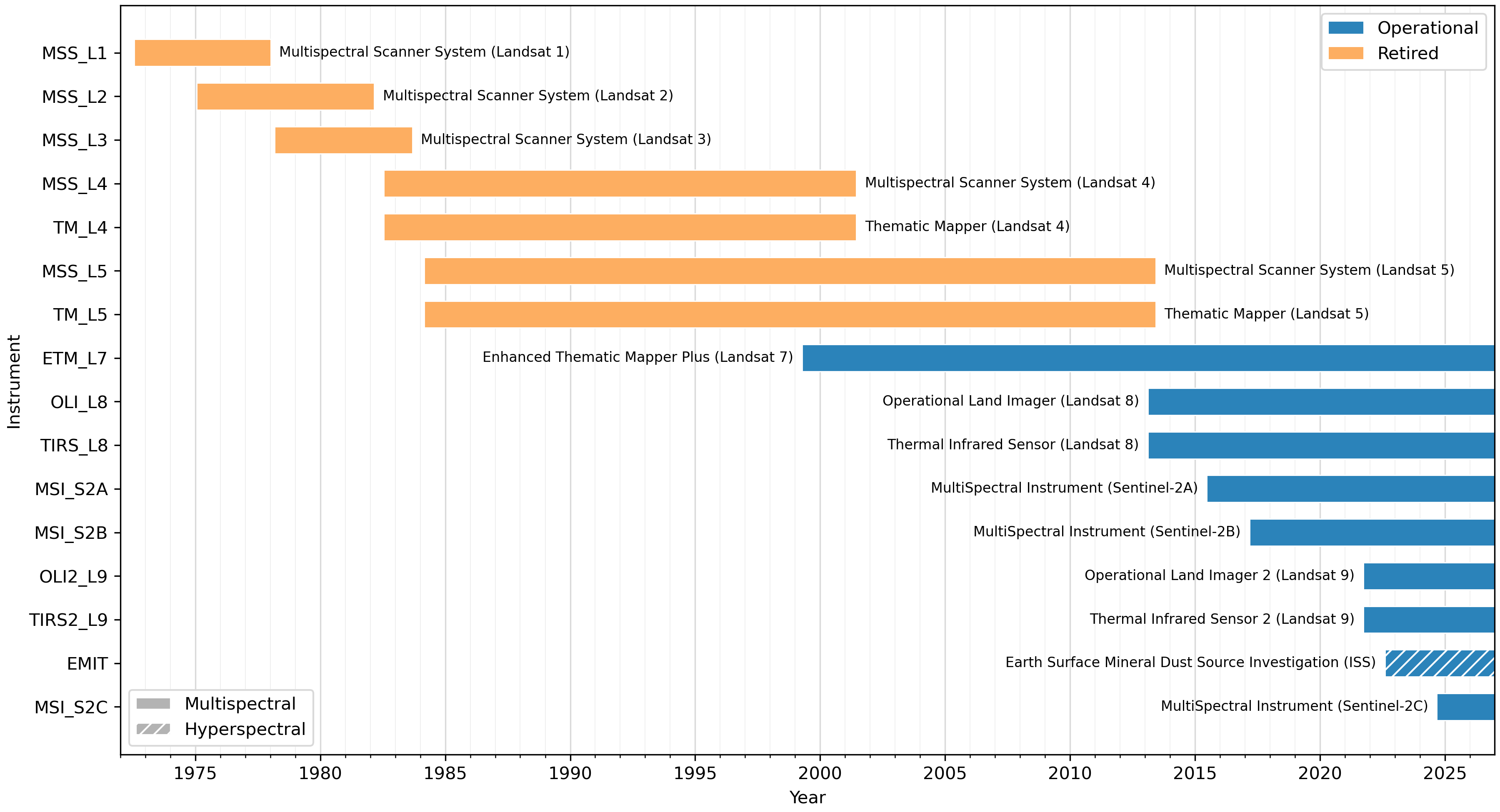}
	\caption{\textbf{Overview of the Earth observation (EO) instruments included in the pilot release of the catalogue.} The metadata used to generate this figure were programmatically extracted from the core schema properties ingested for each instrument.}
\label{fig:timeline}
\end{center}
\end{figure}

\begin{figure}[ht!]
\begin{center}
	\includegraphics[width=1.0\textwidth]{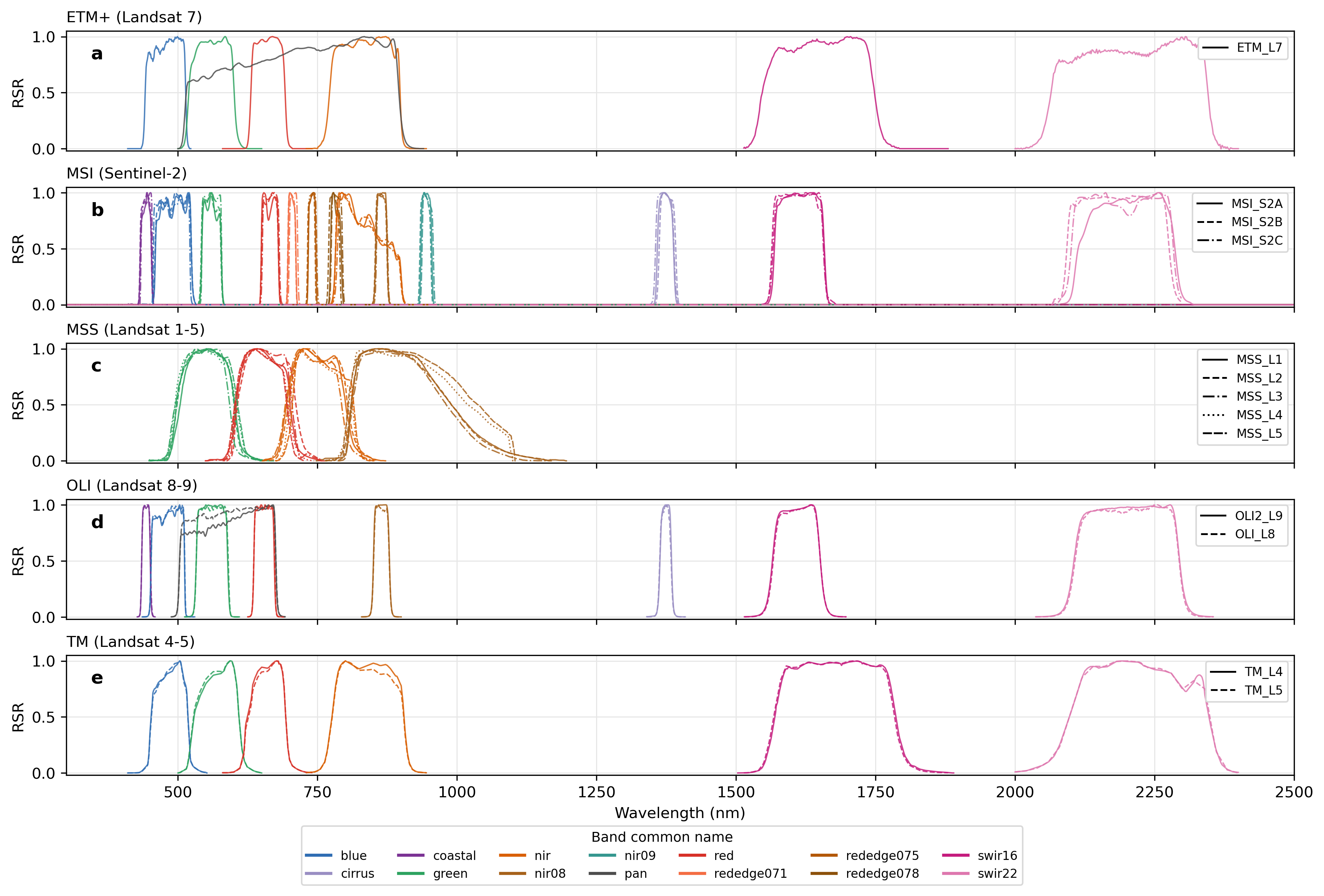}
	\caption{\textbf{Spectral response functions (SRFs) of the optical bands represented in the pilot release of the catalogue.} Curves show the relative spectral response (RSR) of each band for the multispectral instruments included in the pilot release. Band groups follow the \texttt{common\_name} attribute used by the STAC Electro-Optical Extension when available; for MSS instruments, this categorisation was inferred from the band names and spectral configuration.}
\label{fig:srf_optical}
\end{center}
\end{figure}

\begin{figure}[ht!]
\begin{center}
	\includegraphics[width=0.6\textwidth]{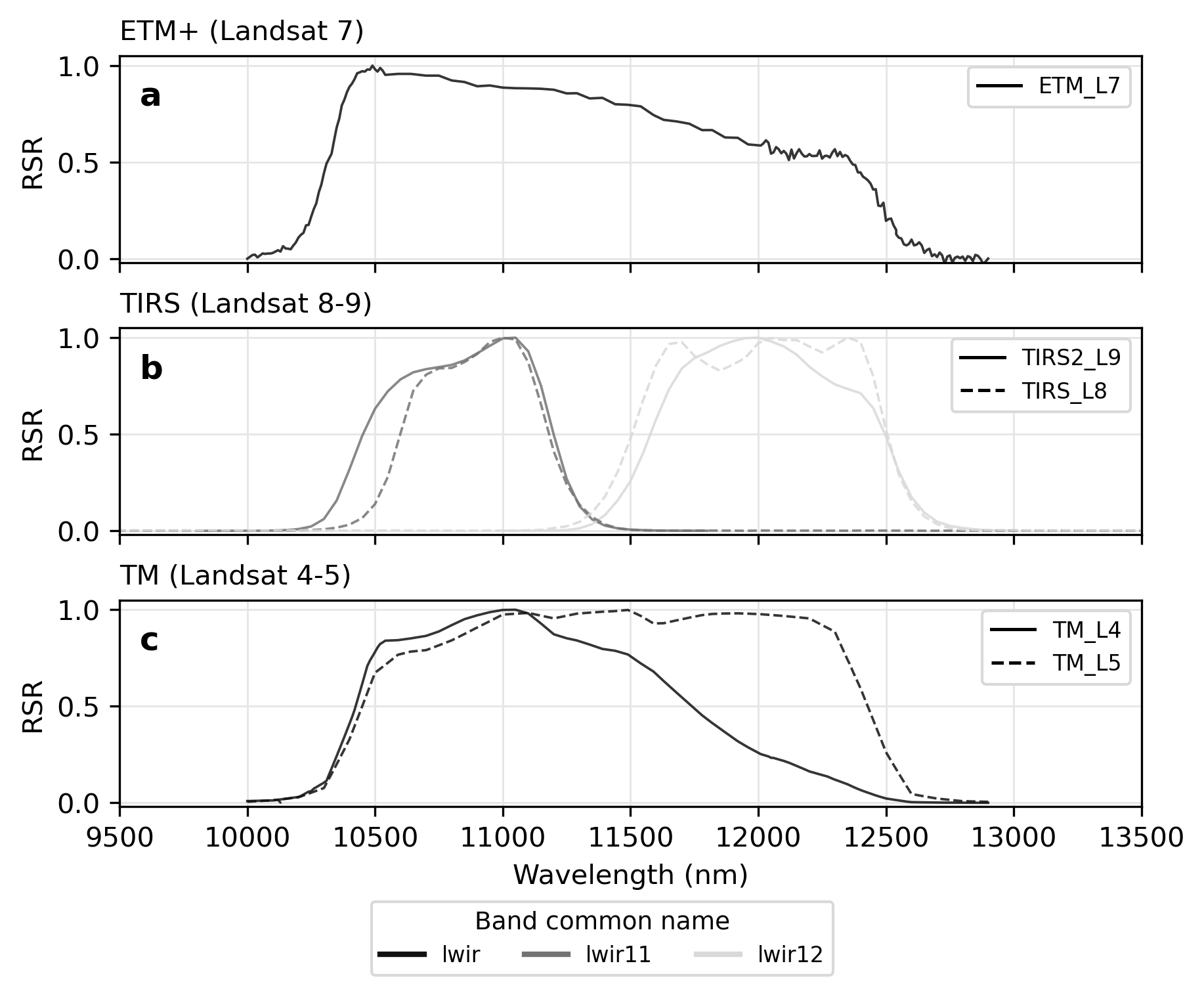}
	\caption{\textbf{Spectral response functions (SRFs) of the thermal bands represented in the pilot release of the catalogue.} Curves show the relative spectral response (RSR) of each thermal band for the multispectral instruments included in the pilot release. Band groups follow the \texttt{common\_name} attribute used by the STAC Electro-Optical Extension when available.}
\label{fig:srf_thermal}
\end{center}
\end{figure}

All pilot instruments were successfully catalogued in AEOI and are summarised in Figure~\ref{fig:timeline}. To further demonstrate the functionality of the generated catalogue, we used the stored spectral metadata to plot the available SRFs in the optical and thermal regions (Figures~\ref{fig:srf_optical}~and~\ref{fig:srf_thermal}, respectively). These examples illustrate that the catalogue can store and expose instrument-level spectral information in a form that is directly useful for EO workflows and analyses. This initial set corresponds to version~v0.1.0 of the catalogue; later versions may include additional instruments.

\section{Discussion}\label{discussion}

This section discusses the potential integration of AEOI with existing open resources, its relevance for AI workflows, possible future extensions, programmatic access, and the limitations inherent to maintaining a free, open, and community-driven catalogue.

\subsection{Potential Integration with Existing Resources}

\subsubsection{Awesome Spectral Indices (ASI):}

Awesome Spectral Indices (ASI) is an ecosystem that provides an open catalogue of spectral indices together with interfaces for Python, Julia, and GEE \citep{montero2023asi}. The ASI catalogue could be enriched by linking each spectral index to 1) the instrument for which the index was originally developed, when applicable, and 2) the instruments in AEOI for which the index can be calculated. Such links would help researchers identify the instrumental context in which a spectral index was defined and determine which other instruments can support its computation.

\subsubsection{Transparent Access to Cloud-Optimized datasets (TACO) Specification:}

The TACO specification\footnote{{https://tacofoundation.github.io}} is a FAIR-compliant, cloud-native specification that defines a formal and scalable format for packaging and sharing AI-ready EO datasets \citep{aybar2026missing,pellicer2026taco}. Datasets created with TACO, such as CloudSEN12+\footnote{{https://huggingface.co/datasets/tacofoundation/cloudsen12}} and MethaneSET\footnote{{https://huggingface.co/datasets/tacofoundation/methaneset}}, could benefit from links to the instruments registered in AEOI. Such links would provide a clearer description of the instruments used to acquire the data in a dataset, helping users identify characteristics relevant to AI model development and potentially use this information directly as model inputs (Section~\ref{ai}).

\subsubsection{SpatioTemporal Assets Catalog (STAC):}

STAC is one of the most widely used specifications for describing and cataloguing EO data. STAC already provides a comprehensive set of attributes for describing datasets and a well-established extension mechanism for representing additional metadata. An AEOI-related STAC extension could link catalogue instrument properties directly to STAC collections, helping users understand the instrument characteristics associated with specific datasets in a manner analogous to the proposed integration with TACO.

\subsection{Artificial Intelligence}\label{ai}

The metadata gathered in AEOI are relevant for AI modelling with EO data. Instrument characteristics can inform the design of model architectures, preprocessing strategies, and downstream task definitions. Moreover, because AEOI provides these characteristics in a standardised machine-readable form, they can be used directly by metadata-aware AI models. Such models could incorporate instrument metadata as prior information or as explicit encodings, enabling more robust sensor-aware or sensor-agnostic approaches \citep[e.g., ][]{francis2022sensei}. In this setting, a single model could potentially operate across datasets from different instruments by conditioning predictions on the relevant instrument metadata.

\subsection{Further Extensions}

One of the main advantages of the proposed schema is its extensibility. New extensions can be introduced to address additional instrument properties, data-access mechanisms, or domain-specific requirements. The current implementation has a strong focus on optical instruments, but the same approach can support extensions for radar and LiDAR systems. Such extensions could describe properties relevant to current radar missions such as Sentinel-1 and ESA Biomass, retired radar missions such as TerraSAR-X and ALOS-PALSAR, spaceborne LiDAR missions such as GEDI, and terrestrial laser-scanning systems.

Additional data-access extensions could also be developed. The existing Earth Engine and Planetary Computer extensions share a common structure that could be adapted to other access points, such as EarthData. In particular, the Planetary Computer extension includes a STAC endpoint property, making it a useful template for other STAC-based catalogues while preserving a standardised representation.

A further extension could represent cross-links to external resources. Because AEOI is not the only resource that records EO instrument characteristics, such an extension could connect AEOI items to equivalent or related entries in other catalogues. For example, the Index DataBase (IDB), which catalogues spectral indices, also includes sensor information. AEOI instruments could also be linked to relevant data-source catalogues such as EarthData.

\subsection{Programmatic Access}

The catalogue was designed to be machine-readable from the outset. This model has proven useful in the ASI ecosystem, and AEOI extends this idea by providing the complete catalogue as a JSON file. This format makes the registry accessible from most programming languages and facilitates the development of APIs for common EO data-processing environments, including Python, R, Julia, and GEE interfaces in JavaScript or Python.

\subsection{Limitations and Future Work}

AEOI was designed to make community contributions straightforward: missing instruments can be added through simple YAML files. This design, however, also implies a maintenance burden. Thousands of EO instruments have been developed, and keeping a registry complete and up to date requires sustained effort. Version~v0.1.0 includes 16 items, and this number is expected to increase in future releases. Community engagement will therefore be essential. A similar pattern occurred in Awesome Spectral Indices, which grew from approximately 60 spectral indices in version~v0.1.0 to more than 270 in version~v0.10.0 through contributions from maintainers and the remote sensing community. We expect AEOI to grow in a similar manner, beginning with widely used instruments and gradually expanding as users request and contribute the instruments needed for their workflows. Future item creation could be accelerated with large language models that retrieve information, draft input files, and iterate against validators. However, such workflows should remain under human supervision, particularly for the audit phase. AI systems could instead support maintainers by preparing draft pull requests for new instruments. As the catalogue grows, an initial Python API is planned to provide programmatic access in one of the most widely used languages for EO data operations.

\section{Conclusions}\label{conclusions}

In this paper, we introduced Awesome Earth Observation Instruments, an open-source, machine-readable, and community-driven catalogue of EO instruments. The catalogue is based on a compact JSON Schema that describes general instrument properties and can be extended through additional schemas for spectral, optical, geometric, and data-access metadata, including Earth Engine and Planetary Computer access points. The catalogue is designed to be extensible by the community while maintaining a rigorous validation workflow that combines human audit and automatic schema validation before ingestion. Community members can contribute missing instruments or propose new extensions. We foresee AEOI becoming a useful integration point for existing resources such as Awesome Spectral Indices and TACO, and a practical reference for EO data users who require instrument characteristics and metadata for reproducible analysis workflows.

\section*{Code and Data Availability}

The Awesome Earth Observation Instruments (AEOI) catalogue and the code used to generate Figures~\ref{fig:timeline}-\ref{fig:srf_thermal} are publicly available on GitHub\footnote{{https://github.com/awesome-spectral-indices/awesome-earth-observation-instruments}}\textsuperscript{,}\footnote{{https://github.com/davemlz/srf-aeoi}}.

\section*{Declaration of AI Use}

We used ChatGPT for language editing and Codex for programming assistance. All content and code were reviewed and edited as needed, and we take full responsibility for the published work.

\section*{Acknowledgments}

All authors acknowledge support by the Climate Change AI Innovation Grants program, hosted by Climate Change AI with the support of Quadrature Climate Foundation, the Global Methane Hub, Google DeepMind, and the Canada Hub of Future Earth. D.M. and M.D.M. acknowledge support from the European Space Agency --- ESA (`cube.ng --- No Gap Earth System Data Cubes`, CCN1 of ARCEME --- Adaptation and Resilience to
Climate Extremes and Multi-hazard Events, ESA Contract No. 4000144482/24/I-KE). L.G.C. and C.A. acknowledge support from the Spanish Ministry of Science, Innovation and Universities (grant PID2023-148485OB-C21/C22 funded by MCIU/AEI/10.13039/501100011033 ERDF, EU).

\bibliographystyle{abbrvnat}
\bibliography{references}

\end{document}